# Photoneutron Yield for an Electron Beam on Tantalum and Erbium Deuteride

Andrew K. Gillespie [a]*, Cuikun Lin [a], and R. V. Duncan [a]

**AFFILIATIONS**

[a]*Department of Physics and Astronomy, Texas Tech University, Lubbock Texas 79409, USA*
*Author to whom correspondence should be addressed: a.gillespie@ttu.edu

**ABSTRACT**

An electron beam may be used to generate bremsstrahlung photons that go on to create photoneutrons within metals. This serves as a low-energy neutron source for irradiation experiments [1-3]. In this article, we present simulation results for optimizing photoneutron yield for a 10-MeV electron beam on tantalum foil and erbium deuteride ($ErD_3$). The thickness of the metal layers was varied. A tantalum foil thickness of 1.5 mm resulted in the most photons reaching the second metal layer. When a second metal layer of $ErD_3$ was included, the photoneutron yield increased with the thickness of the secondary layer. When the electron beam was directly incident upon a layer of $ErD_3$, the photoneutron yield did not differ significantly from the yield when a layer of tantalum was included. The directional photoneutron yield reached a maximum level when the thickness of the $ErD_3$ layer was around 12 cm. About 1 neutron was generated per $10^4$ source electrons. When using a 2-mA beam current, it is possible to generate up to $10^{12}$ neutrons per second, making this combination a relatively-inexpensive neutron generator.

## I. INTRODUCTION

In 1932, James Chadwick announced the discovery of the neutron. Since then, this groundbreaking finding has led to numerous applications across various fields, including the utilization of neutrons in fusion reactors, the geochronological dating of rocks, the development of non-intrusive inspection techniques based on neutron technology, the use of radioisotopes for radiotherapy, medical imaging advancements, and even pain relief techniques. [4]

However, a major obstacle in the further advancement of these applications is often the lack of suitable neutron sources. There are many ways to produce neutrons, including: (1) Alpha neutron sources (2) Gamma neutron sources (3) Spontaneous fission neutron sources such as Cf-252. (4) Fission reactors (5) Accelerators. This also includes D-D (deuterium-deuterium) neutron generators and D-T (deuterium-tritium) neutron generators. **Table 1** provides a summary of the different types of neutron sources along with their respective yields.

**Table 1:** Neutron sources and respective yields

|  | Source | Target | Most Common Yield [5] |
|---|---|---|---|
| Alpha Neutron Sources | Am-241; Pu-238, Pu-239, Po-210; Ra-226 | Be, Li, F, B | AmBe 2.0-2.4x$10^6$ neutrons/sec. per Ci  PuBe, 1.5-2.0 x$10^6$ n/s per Ci |
| Gamma Neutron Source | Gamma emitting core | Be-9; D; mix of Sb-124 and Be-9 | 0.2-0.3 x$10^6$ n/s per Ci |
| Spontaneous Fission Sources | Cf-252 |  | 4.4 x$10^9$ n/s per Ci |
| Fission Reactors | $^{235}$U + n → $^{236}$U → $^{141}$Ba + $^{92}$Kr + 3n |  | $10^{12}$ to $10^{15}$ n/cm$^2$ s – $10^{12}$ n/s per megawatt (MW) |
| Accelerators | betatron, synchrotron and linear accelerators | Be-9; D, Ta, W, Pb | $10^{12}$ n/s |
|  | D | Ti/D, Ti/T | $10^6$-$10^9$ n/s, $10^9$-$10^{10}$ n/s |

Commercially-available neutron generators can cost between $100,000 - $300,000 depending on the model and desired neutron output. [6] Among the various options, nuclear fission reactors undoubtedly provide the highest neutron flux. However, these reactors are often not readily available or easily constructed. The costs associated with building new nuclear units are substantial, with estimates ranging from $5,500/kW to $8,100/kW or between $6 billion and $9 billion for each 1,100 MW plant, considering factors such as escalation and financing costs [7]. While Cf252 is an option for neutron production, it is generally not suitable for long-term usage due to its short half-life and high price, exceeding $2 million per Curie. Overall, a neutron source with a high thermal neutron flux that is economically viable and suitable for installation in industries or clinics is not widely available. Among the remaining options, the use of electron accelerators with low beam energies to produce neutrons through the (γ, n) reaction has garnered significant attention due to the availability and comparably lower cost of such accelerators as listed in **Table 2**.

**Table 2:** Commercially available electron generators

| Manufacture | Electron Beam Accelerator Model [8] | Type | Energy (MeV) |
|---|---|---|---:|
| IBA Industrial Solutions | TT-50 | RF-SCR | 10 |
| | TT-100 | RF-SCR | 10 |
| | TT-200 | RF-SCR | 10 |
| | TT-300 | RF-SCR | 10 |
| | TT-1000 | RF-SCR | 7.5 |
| NIIEFA | UEL-10-D | RF-Linac | 10 |
| NIIEFA | Elektron 23 | DC | 1 |
| BINP | ILU-10 | RF-SCR | 5 |
| BINP | ILU-14 | RF-Linac | 10 |
| BINP | ELV-12 | DC | 1 |
| Varian | Linatron | RF-Linac | 9 |
| Mevex | Linac | RF-Linac | 3 |
| Wasik Assoc. | ICT | DC | 3 |
| Getinge Group | Linac | RF-Linac | 10 |
| Vivirad S.A. | ICT | DC | 5 |

One of the notable highlights on the list is the compact TT50 Rhodotron™. The design objective for this unit is to achieve a beam energy of 10 MeV and a power output of 20 kW using 80 cm cavity diameter unit [9]. With its impressive energy efficiency of 20%, this unit will serve as an excellent tool for small- to medium-sized service providers as well as research and development institutions. FARSHID et al. estimated a neutron flux of $10^{12}$ n/cm$^2$/s, with average energies of 0.9 MeV, 0.4 MeV, and 0.9 MeV for Pb, Ta, and W targets, respectively, using the RHODOTRON TT200 (IBA) accelerator, as detailed in [10]. Chakhlov et al. also reported using 10 MeV electrons to produce bremsstrahlung beams that irradiated LiD, Be, depleted U, and Pb targets to generate photoneutrons [11,12]. **Figure 1** illustrates a typical design of the photoneutron generator. In this setup, when an accelerated electron beam interacts with a Bremsstrahlung target such as Ta, W, or Pb, Bremsstrahlung radiation is emitted. The total energy and yield of photons generated from the target at various thicknesses are significant parameters to take into account in this process [13].

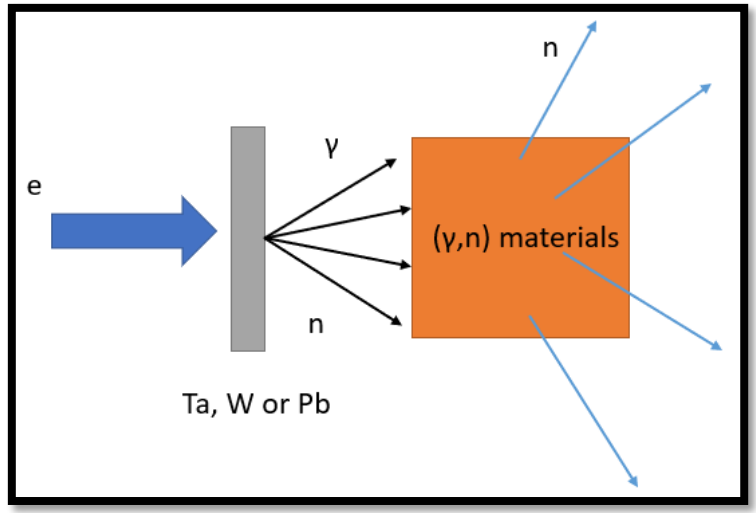

**Figure 1:** A simplified model of a photoneutron generator involving an electron beam incident on tantalum, tungsten, or lead. Photons are created and participate in (γ, n) reactions. [14]

Tsechanski [15] and Berger [16] have demonstrated that when the target thickness exceeds the optimal value, the efficiency decreases due to a greater absorption of braking radiation within the target material. It is important to note that neutrons can also be generated using electrons with an energy of 10 MeV. According to Mahmod et al., by using a 10 MeV electron source and directing it towards a 2 cm-thick Pb material, the estimated photoneutron yield is $1.7 \times 10^{-5}$ photoneutrons per electron. [17]. This implies that for a 10 kW accelerator with a 10 MeV and 1 mA electron source, the neutron production would be approximately $10^{11}$ n/s. This production rate is comparable to D-D neutron generators and D-T neutron generators but comes at considerably higher cost. This is mainly due to the high threshold energies (>6 MeV) required by these high atomic number (Z) materials, resulting in limited photon utilization and lower cross sections. **Table** 3 provides an overview of materials along with their corresponding high threshold reactions for neutron generation.

**Table 3:** Materials and high threashold reactions for neutron generation [10,18]

| Nuclei | Threshold (MeV) | Isotope Abundance (%) |
|---|---|---|
| $^{206}$Pb | 8.09 | 24.1 |
| $^{207}$Pb | 6.74 | 22.1 |
| $^{208}$Pb | 7.37 | 52.4 |
| $^{181}$Ta | 7.58 | 99.9 |
| $^{180}$W | 8.41 | 0.12 |
| $^{182}$W | 8.07 | 26.3 |
| $^{183}$W | 6.19 | 14.3 |
| $^{184}$W | 7.41 | 30.7 |
| $^{186}$W | 7.19 | 28.6 |

Another viable option for photoneutron production is to utilize low threshold materials. **Table 4** provides an overview of materials along with their corresponding low threshold reactions for neutron generation. D (deuterium) and Be-9 (beryllium-9) are particularly noteworthy due to their low binding energies of 2.23 MeV and 1.66 MeV, respectively. In a recent study conducted by a team at NASA-Glenn SFC, a novel d+D fusion process in metals was investigated at low deuteron projectile energy but under high-flux gamma radiation. Steinetz *et al*. [3] performed experimental investigations utilizing an electron beam. The electron beam had an energy range of 0.45 to 3.0 MeV and a current of 10 to 30 mA, which was directed towards a tantalum target. By generating bremsstrahlung X-rays up to 2.9 MeV, the beam irradiated both $ErD_3$ and $TiD_2$ samples, resulting in the production of photo-neutrons. By leveraging these low threshold materials, more efficient and cost-effective photoneutron generation can be achieved.

**Table 4:** Materials and low threshold reactions

| Nuclide | Threshold (MeV) | Reaction |
| --- | --- | --- |
| $^2$D | 2.23 | $^2H(\gamma,n)^1H$ |
| $^6$Li | 3.70 | $^6Li(\gamma,n+p)^4He$ |
| $^6$Li | 5.67 | $^6Li(\gamma,n)^5Li$ |
| $^7$Li | 7.25 | $^7Li(\gamma,n)^6Li$ |
| $^9$Be | 1.67 | $^9Be(\gamma,n)^8Be$ |
| $^{13}$C | 4.90 | $^{13}C(\gamma,n)^{12}C$ |

To further comprehend and optimize the neutron yield, comprehensive simulations were conducted in our paper. The objective was to optimize the photoneutron yield using a 10-MeV electron beam incident on tantalum foil and erbium deuteride. The simulations involved varying the thickness of the metal layers and $ErD_3$ layers in order to explore the impact on the neutron generation process.

## II. METHODS AND SIMULATION GEOMETRY

Three simulation sets were performed using Monte Carlo n-Particle (MCNP™ 6.2) transport calculations [19]. All simulations used the same 10-MeV electron beam as a planar source. All simulations used one hundred million histories in their calculations. Two vacuum spaces were included before and after the metal layers to determine particle counts. The first set of simulations involved a disk layer of tantalum metal with a radius of 2.0 cm. The second set involved a cylinder of erbium deuteride with a radius of 10.0 cm. The third set involved both a layer of tantalum and a layer of erbium. These simulation geometries are shown in **Figure 2**.

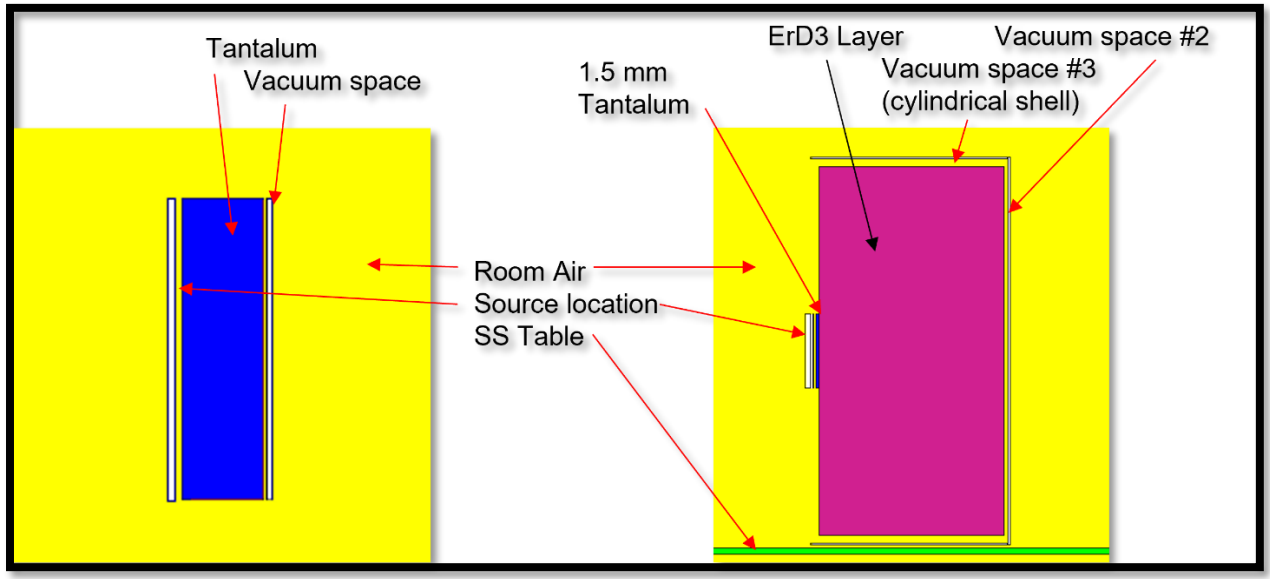

**Figure 2:** *Left*: Simulation set #1 involving only a layer of tantalum. *Right*: Simulation set #3 involving a 1.5 mm-thick layer of tantalum and a layer of erbium deuteride. Simulation set #2 uses this same geometry template as in simulation set #3 with the tantalum layer removed.

For simulations involving the 10-cm radius cylinder of ErD$_3$, an additional tally region was added surrounding the cylindrical wrap. This tally zone, along with the tally region behind the metal layers, allowed for the detection of any particles escaping the layers in specific directions of interest. Though ErD$_3$ has a crystal density near 7.6 g/mL, the synthesis process often involves embrittlement of the metal. When in powder form, ErD$_3$ has a tap density near 3.8 g/mL. Therefore, this 50% packing fraction was used for the ErD$_3$ layer within these simulations.

Within MCNP, the F4 tally calculates the track length estimate of cell flux in units of [1/cm$^2$]. This is calculated by multiplying the importance weight by the track length and is normalized by the cell volume. When multiplied by the source strength in particles per second, this results in the average flux across the tally zone in units of [particles/s/cm$^2$]. The F4 tally can be useful in estimating the neutron yield reaching a volume of interest. The F4 tally may also be split into energy bins to obtain an energy spectrum of particles entering the zone. MCNP is also able to output particle tracks entering each defined volume cell.

### III. RESULTS AND DISCUSSIONS

To investigate the number of electrons transmitted through the tantalum layer, particles entering the first and second tally regions were compared. The percent of electrons reaching the tally region behind the tantalum is shown in **Figure 3**.

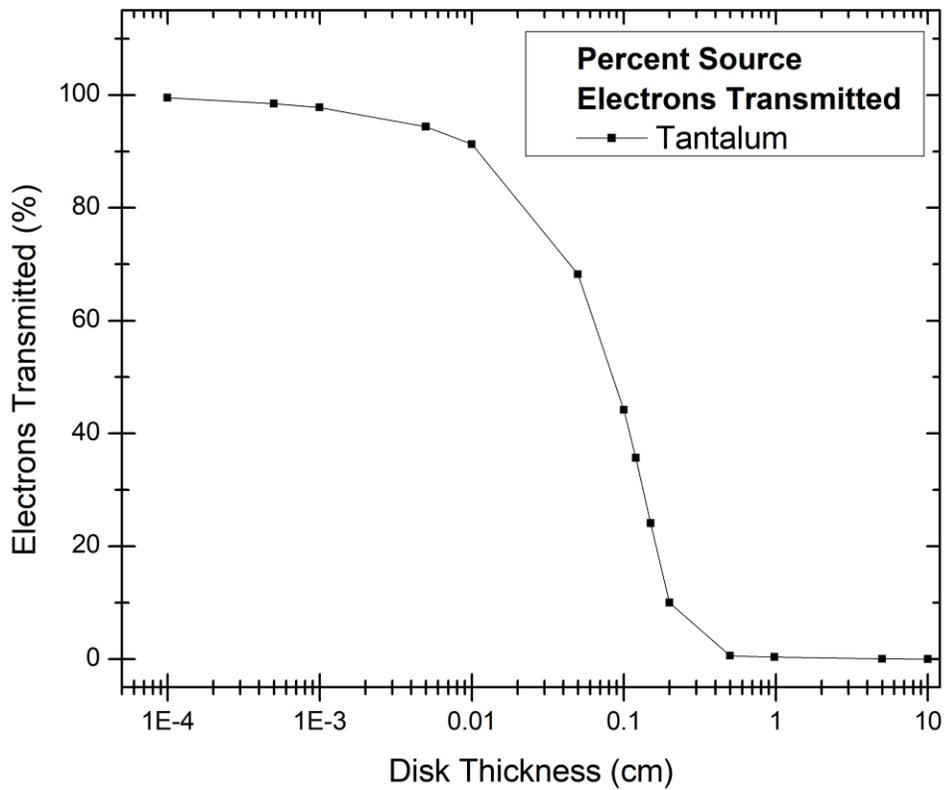

**Figure 3:** The percentage of electrons reaching the tally region behind the tantalum layer as the thickness of the layer is varied.

For layers < 1 µm, almost all of the 10-MeV electrons are transmitted through the tantalum to the second tally region. For thicker layers > 0.5 cm, electrons interact significantly and almost zero electrons are transmitted. When the tantalum layer is still thin, but only about 50% of electrons are transmitted, the deccelerating electrons create bremsstrahlung radiation that is able to escape the tantalum layer. There should be an optimal tantalum thickness that permits the maximum amount of bremsstrahlung photons to escape and continue on to generate photoneutrons. **Figure 4** displays the number of photons created as a function of tantalum thickness.

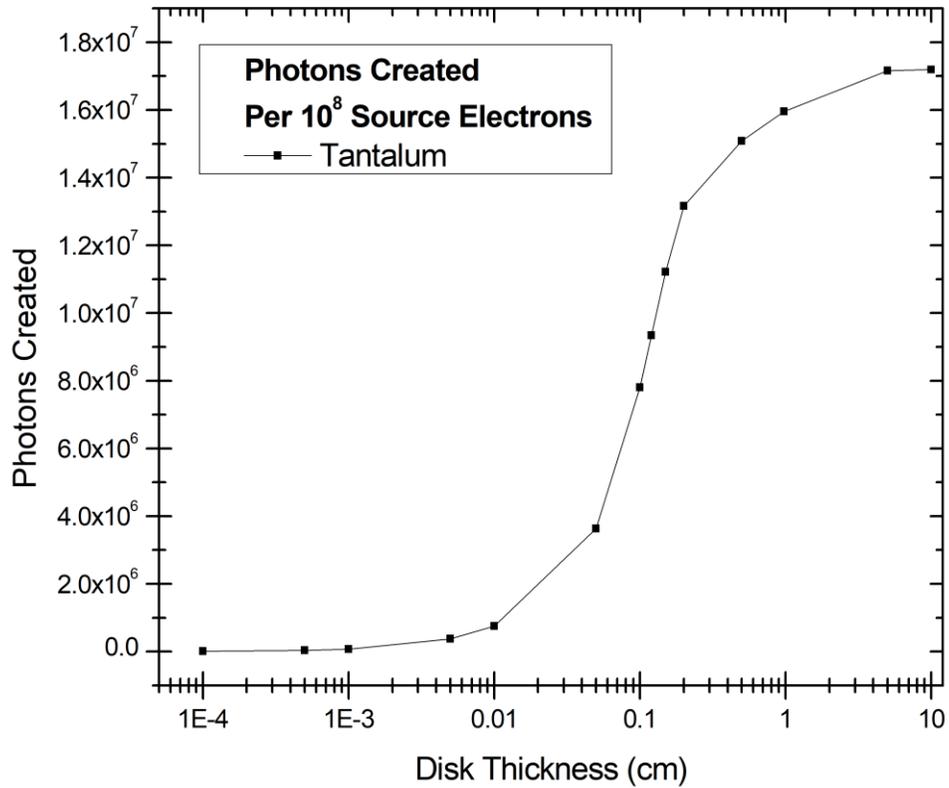

**Figure 4:** The number of photons created as the thickness of the tantalum layer is varied. The total number of photons created increases with tantalum thickness.

The number of photons created increases with thickness of the tantalum layer. However, only the photons that escape the layer will go on to generate photoneutrons in a secondary metal layer. It is more enlightening to investigate the photon tallies in a region immediately behind the tantalum layer. **Figure 5** displays the F4 photon tally within the vacuum region immediately behind the tantalum

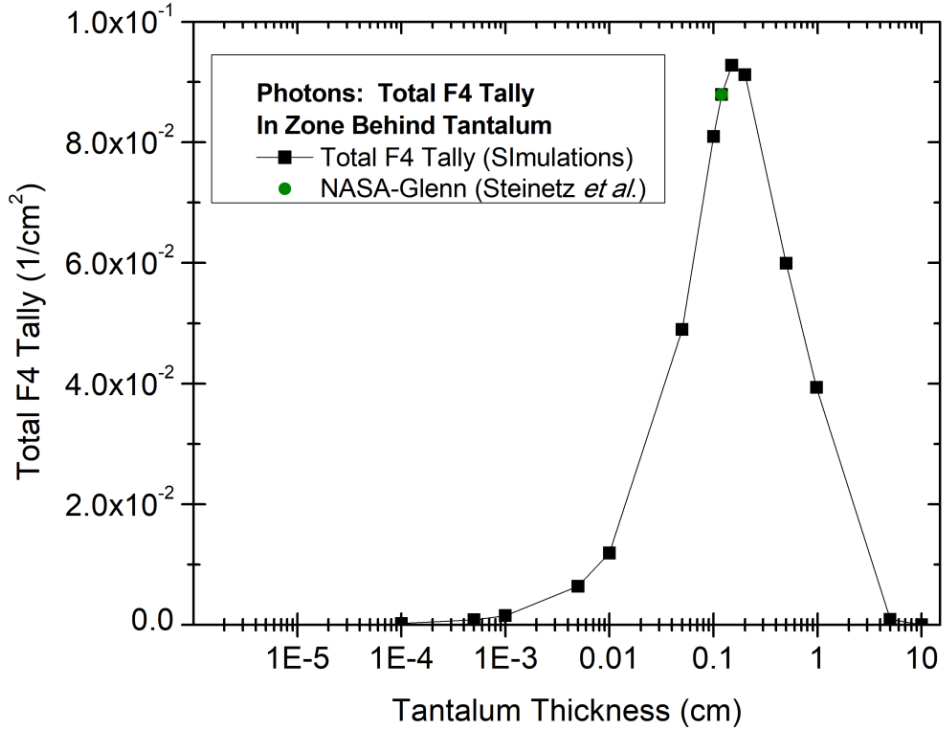

**Figure 5:** The average track length estimate for photons reaching the tally region behind the tantalum layer as the thickness of the layer is varied. When the source strength, in units of [particles/s], is multiplied by this tally, one obtains the average flux in units of [particles/s/cm$^2$].

A tantalum foil thickness of 1.5 mm resulted in the most photons reaching the second metal layer. Steinetz et al. used a tantalum foil with a thickness of 1.2 mm, which yields a high photon tally very close to the peak in this series of simulations. The energy spectrum of these photons is also moderated with increasing tantalum thickness. As more electrons interact with the layer, more photons are created. However, with increased layer thickness, fewer of those photons are able to escape and reach the tally zone. **Figure 6** displays the energy spectrum of the photons reaching the tally zone.

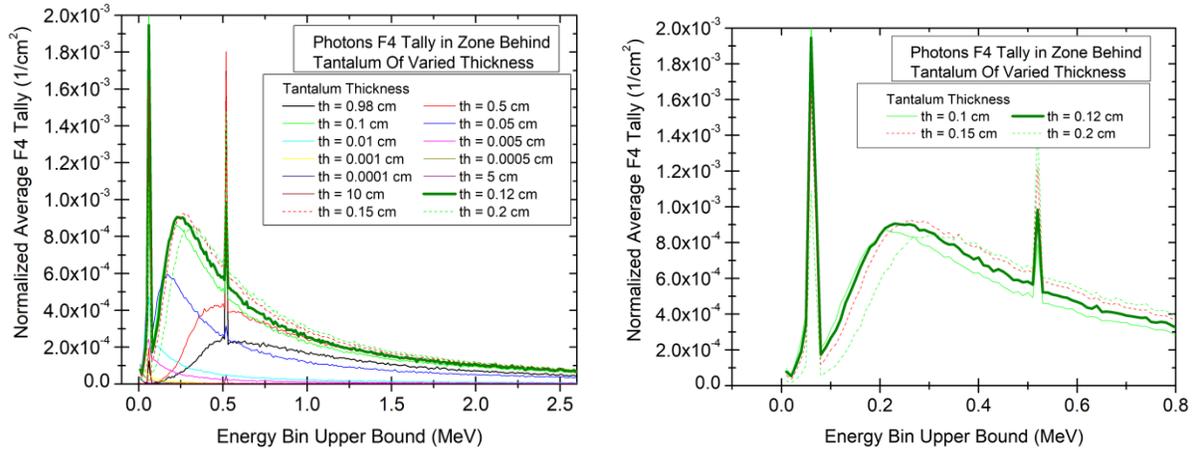

**Figure 6:** The bremsstrahlung photon energy spectrum with varied tantalum thickness.

For a tantalum foil with a thickness near 1.5 mm, prominent peaks around 60 keV, between 230 – 310 keV, and around 520 keV are observed. These photon spectra are comparable to those calculated by Huang *et al.* which shows a similar peak around 0.5 MeV.[20] Maximizing the number of photons reaching the secondary metal layer will result in the most photoneutrons being generated. The tantalum foil thickness was set to 1.5 mm and a secondary layer of $ErD_3$ was added to the system. **Figure 7** displays the number of neutrons generated in these layers as a function of the $ErD_3$ layer thickness.

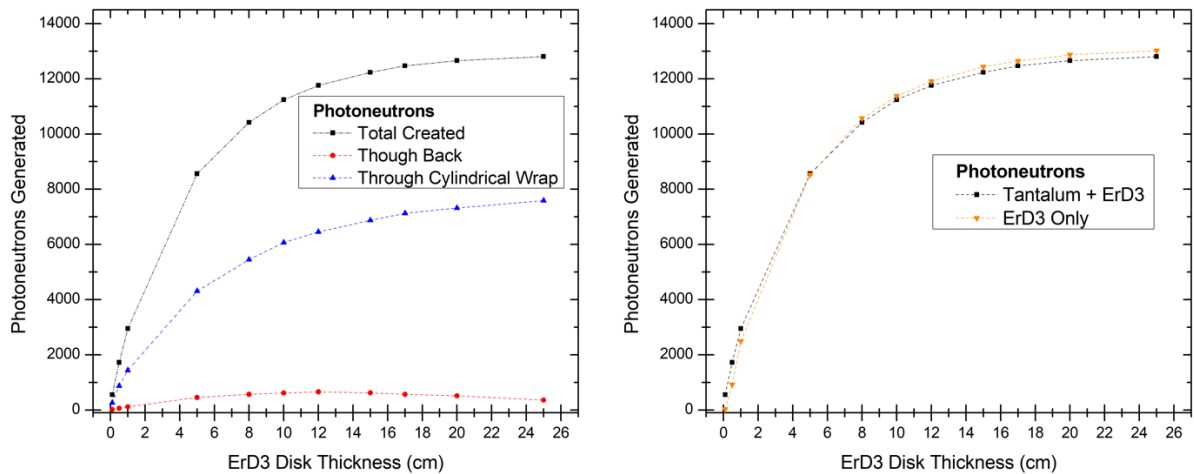

**Figure 7:** *Left*: The photoneutrons generated as a function of increasing $ErD_3$ thickness. *Right*: The total number of photoneutrons created when the electron beam is incident on tantalum and erbium deuteride versus when the beam is incident only upon the erbium deuteride.

When a second metal layer of $ErD_3$ was included, the photoneutron yield increased with the thickness of the secondary layer. When the electron beam was directly incident upon a layer of $ErD_3$, the photoneutron yield was not significantly different from the yield when a layer of tantalum was included. Therefore, it is not beneficial to include the tantalum layer, and instead, run the electron beam directly into the ErD$_3$ layer to generate the most neutrons.

Many applications of neutron generators are interested in maximizing the neutron flux in a specific direction. In this case, it is important to focus only on the number of neutrons reaching the tally zone behind the ErD$_3$ layer, shown in **Figure 8**.

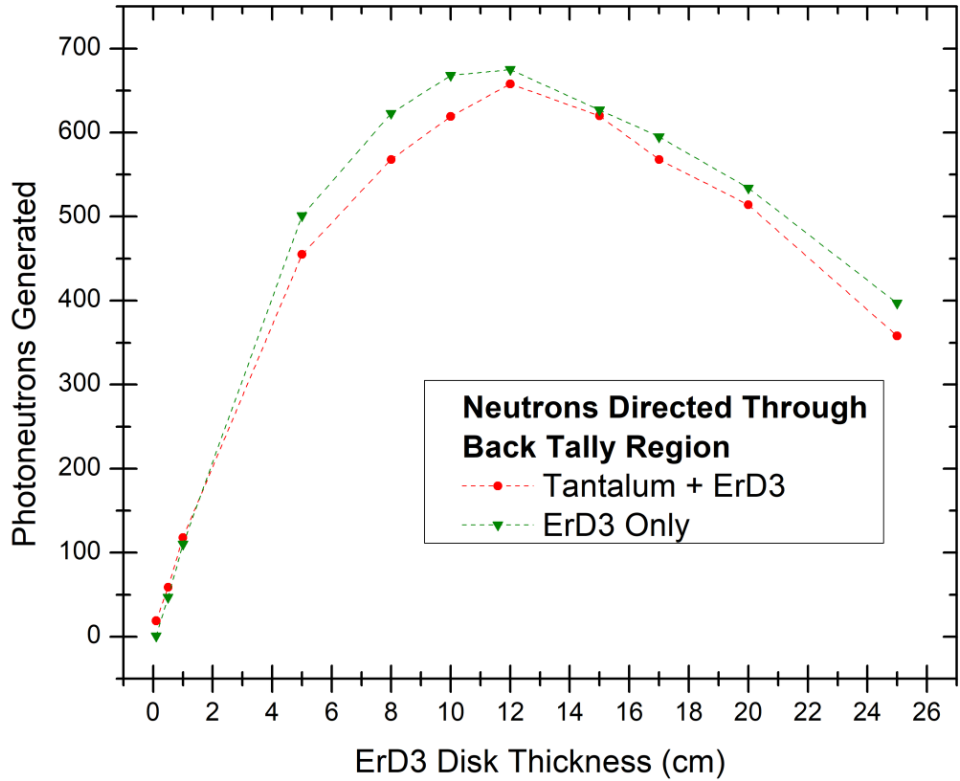

**Figure 8:** The photoneutrons entering the tally zone behind the ErD$_3$ as a function of increasing ErD$_3$ thickness. Results are shown for cases when the electron beam is incident on tantalum and erbium deuteride versus when the beam is incident only upon the erbium deuteride.

A smaller fraction of photoneutrons exit the layer along the same axis as the source. This reaches a peak when the thickness of the ErD$_3$ layer is around 12 cm. A thicker layer results in a higher amount of neutrons being generated overall, but the amount that exit through the back of the ErD$_3$ layer decreases with increasing layer thickness. As shown in **Figure 9**, The average energy of the photoneutrons escaping the layers did not change significantly with layer thickness.

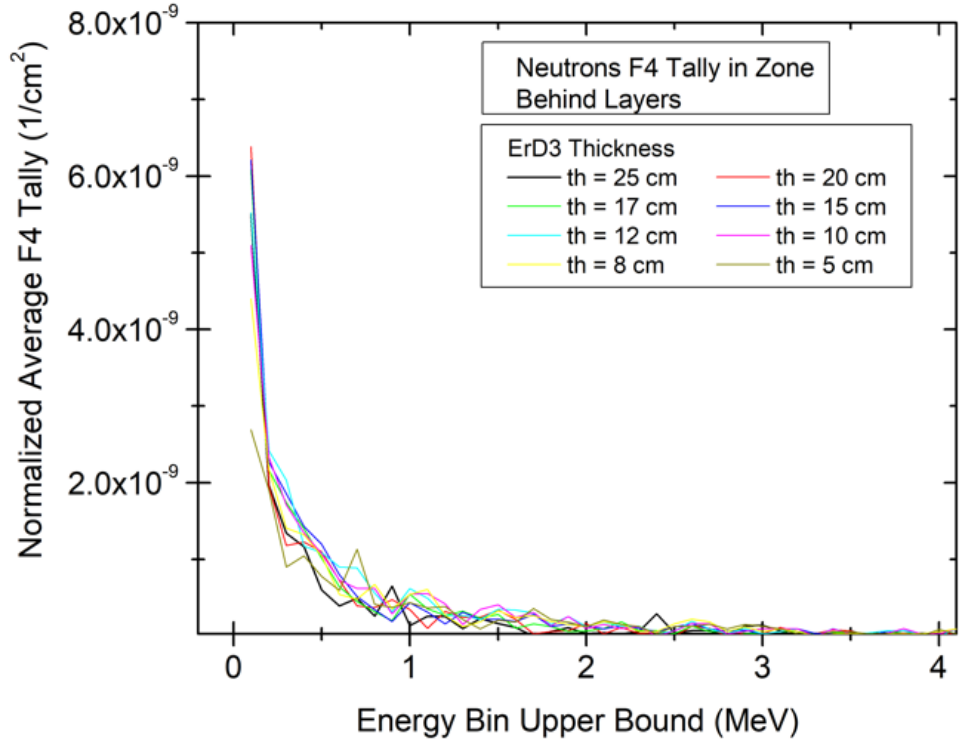

**Figure 9:** The energy spectrum of photoneutrons entering the tally zone behind the $ErD_3$ layer of 12-cm thickness.

The weighted average energy of these directional photoneutrons is near 14 keV. Across all simulations, nearly all neutrons were generated from photodissociation and only between 0.02 – 0.03% were created from (n, xn) reactions within the metals.

All simulations used one hundred million histories in their calculations. Overall, about 1 neutron is generated per $10^4$ source electrons in the beam. Only about 1 directional neutron is generated per $10^6$ source electrons in the beam. A Rhodotron TT50 operates with a 2 mA beam current and 20 kW producing 10-MeV electrons. Since one amp is $6.28 \times 10^{18}$ electrons per second, this 2-mA source should generate about $1.2 \times 10^{16}$ electrons per second. Therefore, this setup would generate about $1.2 \times 10^{12}$ neutrons per second overall and $1.2 \times 10^{10}$ directional neutrons per second.

This demonstrates the feasibility of using an electron beam and bremsstrahlung photons, to generate photoneutrons. Depending on the model specifications, electron generators may be run with low power consumption. When using a 2-mA beam current, it is possible to generate up to around $10^{12}$ neutrons per second, making this combination a relatively-inexpensive neutron generator.

## IV. CONCLUSIONS

The bremsstrahlung photon and photoneutron yield was simulated for a 10-MeV electron beam incident on tantalum foil and erbium deuteride. A tantalum foil thickness of 1.5 mm resulted in the most photons reaching the second metal layer. The directional photoneutron yield reached a maximum level when a 12 cm-thick layer of erbium deuteride was included behind the tantalum layer. When using a 2-mA beam current, it is possible to generate up to $10^{12}$ neutrons per second, making this combination a relatively-inexpensive neutron generator that could be employed in numerous applications, including fusion research.

## V. ACKNOWLEDGEMENTS

The authors would like to thank Dr John Gahl at the University of Missouri for their useful discussions. This work was supported by NIAC Phase I contract No. 80NSSC23K0592 and the Texas Research Incentive Program, and by Texas Tech University. The identification of commercial products, contractors, and suppliers within this article are for informational purposes only, and do not imply endorsement by Texas Tech University, their associates, or their collaborators.## VI. DATA AVAILABILITY

The data that support the findings of this study are available from the corresponding author upon reasonable request.